\documentclass[aps,prc,preprint,tightenlines,showpacs,%
amssymb,byrevtex,nofootinbib]{revtex4}  

\usepackage{epsfig,bm,dcolumn}
\newcommand{\fslash}[1]{\ooalign{\hfil/\hfil\crcr$#1$}}
\begin{document}



\title{Anti-charmed pentaquark $\bm{\Theta_c (3099)}$ from QCD sum rules}


\author{Hungchong Kim}%
\email{hung@phya.yonsei.ac.kr}

\author{Su Houng Lee}%
\email{suhoung@phya.yonsei.ac.kr}

\affiliation{Institute of Physics and Applied Physics,
Yonsei University, Seoul 120-749, Korea}

\author{Yongseok Oh}%
\email{yoh@physast.uga.edu}

\affiliation{Department of Physics and Astronomy,
University of Georgia, Athens, Georgia 30602, U.S.A.}



\begin{abstract}

We construct QCD sum rules for the anti-charmed pentaquark
$\Theta_c (3099)$, recently reported at HERA.
The sum rules are constructed similarly with the $\Theta^+ (1540)$ sum
rules using the anti-charmed analogue of the $\Theta^+$ interpolating field.
The strange quark and quark-gluon mixed condensates, which were important in
the $\Theta^+$ sum rules, are replaced by the gluon
condensates whose contribution to the OPE
is suppressed due to the heavy quark mass.
Our result suggests that the parity of $\Theta_c$ is positive.
We identify the difference from the $\Theta^+$ sum rule, which leads to the
positive parity in this heavy-light pentaquark system.
The obtained mass is similar to the experimental value. 

\end{abstract}


\maketitle

\section{Introduction}

The exotic $\Theta^+ (1540)$ baryon, after its first discovery
\cite{LEPS03} and confirmations in subsequent 
experiments \cite{OTHERS},
has brought huge excitements in hadron physics.
It is a narrow resonance containing 5 quarks, $uudd{\bar s}$, that has not
been observed before.
Though not confirmed, it is believed to be an isoscalar with spin $1/2$,
forming a flavor $\overline{\bf 10}$. 
NA49 Collaboration \cite{Alt:2003vb} later reported the observation of the
narrow resonance, $\Xi^{--} (1862)$, which could be another member of the
same multiplet $\overline{\bf 10}$.
The observation of the anti-charmed pentaquark $\Theta_c^0 (3099)$ in the
$D^* p$ invariance mass spectrum has been recently reported  by H1
collaboration at HERA \cite{Aktas:2004qf}. 
It is anti-charmed analogue of the $\Theta^+ (1540)$ and has the quark
content $uudd{\bar c}$.
The pentaquark with one heavy anti-quark was first studied in
Ref.~\cite{Lip87-GSR87} in a quark model.
Then it has been studied in quark models \cite{Stan98} and Skyrme models
\cite{RS93,OPM94} and attracts recent interests~\cite{CBSC} motivated by
the current experimental activities.
Thus, the pentaquarks are becoming a solid member in hadron spectroscopy
and await a systematic compilation of their properties.

One interesting model for the pentaquark is the diquark-diquark-antiquark
picture of Jaffe and Wilczek (JW) \cite{JW03}.
In this picture, $\Theta^+ (1540)$ is composed by the constituent quarks,
$ud$-$ud$-$\bar {s}$. 
The two diquarks $ud$-$ud$ are identical and the boson symmetry restricts
them to be in a relative $P$-wave.
Thus, the two diquarks combined with an $S$-wave antiquark lead to 
even parity for $\Theta^+$.
The positive parity is also supported by the triquark-diquark picture
\cite{KL03a,Kochelev:2004nd}, the soliton model prediction \cite{DPP97},
the quark potential model calculations \cite{CCKN03b} and quenched lattice
calculation \cite{Chiu:2004gg}.
However, dynamical calculations based on QCD sum rules \cite{SDO03} or lattice
calculations of Refs.~\cite{Sasaki03,CFKK03} support the negative parity
of $\Theta^+$.
Therefore, the $\Theta^+$ parity is an important issue to be settled and
should be determined eventually from reaction mechanisms.
Indeed, various reactions have been suggested to determine
the $\Theta^+$ parity.
Several proposals are based on the order of magnitude of cross sections
and polarization observables in the reactions including
$\gamma N \to K \Theta^+$ \cite{Oh:2003kw,ZANA}, 
$K^+ p \to \pi^+ K^+ n$ \cite{HHO03},
$\gamma n \to K^- K^+ n$ \cite{NT03},
$p+ p \to \Sigma + \Theta$ \cite{ppst} and
$\gamma N \to {\bar K}^* \Theta^+$~\cite{Oh:2003xg}.

Among various QCD sum rule calculations for $\Theta^+$
\cite{Zhu03,MNNRL03,SDO03,Eidemuller:2004ra}, the approach 
proposed by Sugiyama, Doi and Oka
(SDO) \cite{SDO03} is particularly interesting.
Here, the interpolating field for $\Theta^+$ is constructed by mostly
following the JW picture except for the fact that all the quarks are placed
locally. 
The same current has been used in the lattice calculation \cite{Sasaki03}. 
Since QCD sum rules deal with {\it current quarks}, one can certainly form
two different diquarks (with opposite parity) from the $ud$ system and the
boson symmetry no longer applies to the two-diquark system.
Nevertheless, the two-diquark system still has odd parity as in the JW picture.
The QCD sum rule of SDO has some interesting features.    
In the operator product expansion (OPE), each diquark propagates only to the
diquark and the two diquarks with different parities do not mix each other.
This feature is quite welcomed because the diquarks are expected to be a
tightly bound system and hence they may not be easy to diffuse to the others
in their propagations.
The $\Theta^+$ properties in the SDO sum rule are mainly determined by the
nonperturbative effects coming from the anti-strange quark.
In particular, the negative parity, which is opposite to the JW prediction,
is mainly driven by the quark-gluon mixed condensate,
$\langle \bar {s} g_s \sigma \cdot G s \rangle$, which is proportional
to the average quark virtuality in the QCD vacuum.
When the quark mass becomes heavier (like in the constituent quark picture), 
such a virtuality should become smaller and it may be possible to flip the
parity.

Thus, the extension to the charmed analogue $\Theta_c (3099)$ provides
an interesting test for the SDO sum rule and lattice calculations
\cite{Sasaki03}. 
Here, the charm quark is quite heavy so that the constituent-quark 
picture may fit well and the JW prediction for the parity is expected to
be reproduced from QCD. 
In fact, quenched lattice calculation finds the parity of
$\Theta_c (3099)$ to be positive \cite{Chiu:2004uh}.
In the extension to the $\Theta_c (3099)$ sum rules, there are two 
important aspects, which make this sum rule different from the SDO sum rule. 
First of all, since the charm quark is too heavy to form quark condensate,
it gives non-perturbative effects only by radiating gluons.
The quark-gluon mixed condensate 
$\langle \bar {s} g_s \sigma \cdot G s \rangle$, which was the important 
contribution in the $\Theta^+$ sum rule, is replaced by gluonic operators
in the heavy quark expansion that are normally suppressed.
Secondly, the charm quark mass has to be kept finite in the OPE, which can
be done by using the momentum space expression for the charm-quark propagator.
This is different from the light-quark sum rule where the calculation is
performed in the coordinate space and all the quark propagators are obtained
based on the expansion with the small quark mass.
Keeping these two aspects in mind, we construct QCD sum rules for
$\Theta_c (3099)$ and see how they are different from the $\Theta^+ (1540)$
sum rule.

This paper is organized as follows.
In Section II, we introduce the interpolating field for the $\Theta_c$ and
show that it transforms properly under parity.
Section III gives the phenomenological side and Section IV gives the OPE side.
The QCD sum rules for $\Theta_c$ and their analysis are given in Section V.

\section{Interpolating field for $\bm{\Theta_c}$}

In our sum rules, we use the following interpolating field for $\Theta_c$,
\begin{eqnarray}
\Theta_c & = & \epsilon^{abk} 
(\epsilon^{aef} u^T_e C \gamma_5 d_f)(
\epsilon^{bgh} u^T_g C d_h) \Gamma C\bar{c}^T_k\ .
\label{current} 
\end{eqnarray}
Here roman indices $a,b, \dots$ are color indices,  $C$ denotes
charge conjugation, $T$ transpose.
Note that we have introduced the $4\times 4$ matrix $\Gamma$ in front of 
the antiquark, which is to be determined from parity consideration below.
The $C\bar{c}^T$ satisfies the charge-conjugated Dirac equation and
represents the anti-charm quark.
The diquarks, $\epsilon^{aef} u^T_e C \gamma_5 d_f$ and 
$\epsilon^{bgh} u^T_g C d_h$, are isoscalar with spin 0.
Due to the $\gamma_5$ difference, the two diquarks have opposite parities.
To make a local interpolating field, all the quarks are defined at the same
space-time.

To determine $\Gamma$, we consider the parity transformation of $\Theta_c$,
\begin{eqnarray}
\Theta_c' (x') = \gamma_0 \Theta_c (x),~~~x'=(t,-{\bf x})\ .
\label{parity}
\end{eqnarray}
This parity transformation must be recovered when each quark (and the 
antiquark) in $\Theta_c$ transforms similarly, namely
\begin{eqnarray}
q'(x') &=& \gamma_0 q(x),~~ (q=u,d,c)\ .
\label{qtran}
\end{eqnarray}
This constraint in fact leads to the usual nucleon interpolating field
\cite{Griegel:rn} commonly used in nucleon QCD sum rules. 
Under this quark transformation, the two diquarks transform 
\begin{eqnarray}
u'^T (x') C  d'(x') &=& - u^T (x) C d (x)\ ,
\nonumber \\
u'^T (x') C \gamma_5 d' (x') &=&  u^T (x) C \gamma_5 d (x)\ .
\label{diquark}
\end{eqnarray}
The anti-quark is transformed accordingly as
\begin{eqnarray}
C\bar{c'}^T (x') = -\gamma_0 C \bar{c}^T (x)\ .
\label{stran1}
\end{eqnarray}
Substituting these into the interpolating field Eq.~(\ref{current})
and demanding the consistency with Eq.~(\ref{parity}), we find $\Gamma=1$.
This type of interpolating field with $c \rightarrow s$
has been used to investigate the properties of 
$\Theta^+ (1540)$~\cite{Chiu:2004gg,SDO03,Sasaki03}.

\section{Phenomenological side}

QCD sum rules for the $\Theta_c$ are constructed from the following
correlation function,
\begin{eqnarray}
\Pi(q)=i \int d^4 x e^{iq\cdot x} \langle 0 | T (\Theta_c (x), 
{\bar \Theta_c }(0) |0 \rangle \ ,
\label{corr}
\end{eqnarray}
where Eq. (\ref{current}) with $\Gamma=1$ is used as the interpolating field. 
To construct the phenomenological side, we note that the $\Theta_c$
interpolating field can couple to both parities
\cite{Jido:1996ia,Lee:2002jb,SDO03}. 
For the positive parity state, the interpolating field couples through
\begin{eqnarray}
\langle 0 | \Theta_c (x) | \Theta_c ({\bf p}) :P = + \rangle  =
\lambda_+ U_\Theta ({\bf p}) e^{-i p\cdot x}\ ,
\end{eqnarray}
while for the negative parity, it couples through
\begin{eqnarray}
\langle 0 | \Theta_c (x) | \Theta_c ({\bf p}) :P= - \rangle  =
\lambda_- \gamma_5 U_\Theta ({\bf p}) e^{-i p\cdot x}\ .
\end{eqnarray} 
Here, $\lambda_{\pm}$ denotes the coupling strength between the
interpolating field and the physical state with the specified parity.
Using this, we obtain the phenomenological side of Eq. (\ref{corr})
separated into chiral even and odd parts,
\begin{eqnarray}
\Pi^{\rm phen} (q) = - |\lambda_{\pm}|^2 ~ { \fslash{q} \pm m_{\Theta} \over
q^2 -m^2_{\Theta} } +\cdot \cdot \cdot
\equiv \fslash{q} \Pi^{\rm phen}_q + \Pi^{\rm phen}_1 \ , 
\end{eqnarray}
where the plus (minus) sign in front of $m_{\Theta}$ is for positive
(negative) parity. 
The dots denote higher resonance contributions that should be 
parametrized according to QCD duality.
It should be noted however that higher resonances with different parities
contribute differently to the chiral-even and chiral odd 
parts \cite{Jin:1997pb}. Thus, $\Pi^{\rm phen}_q$ and $\Pi^{\rm phen}_1$
constitute  separate sum rules.

The spectral density is given by
\begin{eqnarray}
{1\over \pi} {\rm Im} \Pi^{\rm phen} (q) &=& \fslash{q}  
|\lambda_{\pm}|^2  \delta (q^2 -m^2_{\Theta})
      \pm m_\Theta |\lambda_{\pm}|^2 \delta (q^2 -m^2_{\Theta})
+\cdot \cdot \cdot\ .
\end{eqnarray}
We notice that the chiral-odd part  has opposite sign depending on the
parity while the chiral even part has positive-definite
coefficient.
Thus, the chiral-odd part from the OPE side can determine the parity.

\section{OPE side}

\begin{figure}[t]
\centering
\epsfig{file=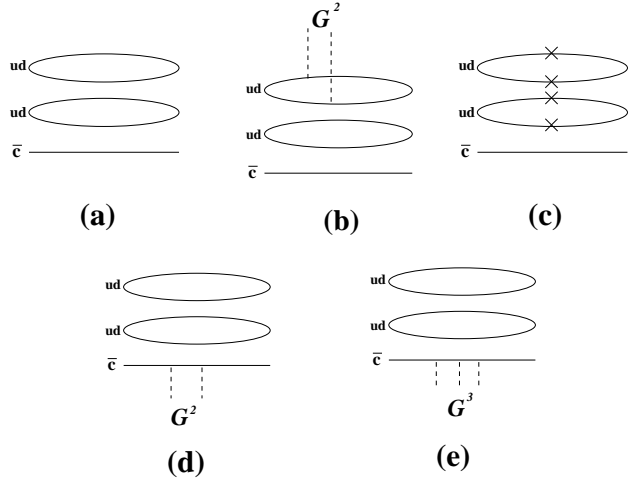, width=0.5\hsize}
\caption{Schematic OPE diagrams representing the two diquarks and
anti-charm quark propagating from 0 to $x$. The solid lines denote 
quark (or anti-charm quark) propagators and the dashed lines
are for gluon. The crosses in (c) denote the quark condensate.
(a) is perturbative contribution
where the diquarks and anti-charm quark just propagate through,
(b) is gluon contribution coming from the diquarks,
(c) is the $\langle \bar{q}q\rangle^4$ contribution from the light quarks,
and (d) and (e) are the gluon contributions from  the charm quark.}
\label{fig1}
\end{figure}

In the OPE side, we calculate the five diagrams shown in Fig.~\ref{fig1}.
To keep the charm quark mass finite, we use the momentum-space expression
for the charm quark propagators.
For the light quark part of the correlation function, we calculate in the
coordinate-space, which is then Fourier-transformed to the momentum space
in $D$-dimension.
The resulting light-quark part is combined with the charm-quark part before
it is dimensionally regularized at $D=4$.

Our OPE is given by
\begin{eqnarray}
\Pi^{\rm ope} (q)=\Pi^{(a)}+\Pi^{(b)}+\Pi^{(c)}
+\Pi^{(d)}+\Pi^{(e)}
\end{eqnarray}
corresponding to each diagram in Fig.~\ref{fig1}.
The imaginary part of each diagram is calculated as
\begin{eqnarray}
{1\over \pi} {\rm Im} \Pi^{(a)} (q^2) &=&  
-{1\over 5\cdot 5!\ 2^{12} \pi^8}  
\int^{\Lambda}_0 du [\fslash{q} (1-u) + m_c ]
\left ( -u q^2 + {m_c^2 u \over 1-u} \right )^5
\ ,\nonumber \\
{1\over \pi} {\rm Im} \Pi^{(b)} (q^2) &=&  
-{1\over 3!\ 3!\ 2^{10} \pi^6} \left \langle {\alpha_s \over \pi} G^2
\right \rangle 
\int^{\Lambda}_0 du [\fslash{q} (1-u) + m_c ]
\left ( -u q^2 + {m_c^2 u\over 1-u} \right )^3
\ ,\nonumber \\
{1\over \pi} {\rm Im} \Pi^{(c)} (q^2) &=&  
-{1\over 54} \langle \bar {q} q \rangle^4 
(\fslash{q}  + m_c )~ \delta (q^2 - m_c^2)
\ ,\nonumber \\
{1\over \pi} {\rm Im} \Pi^{(d)} (q^2) &=&  
-{1\over 5!\ 3!\ 3\cdot 2^{10} \pi^6} \left \langle {\alpha_s \over \pi} G^2
\right \rangle 
\int^{\Lambda}_0 du \left ( u\over 1-u \right )^3
\nonumber \\
&\times & [3 m_c^2 \fslash{q} (1-u) + m_c (1-u)(3-5u)q^2+2u m_c^3 ]
\left ( -u q^2 + {m_c^2 u\over 1-u} \right )^2
\ ,\nonumber \\
{1\over \pi} {\rm Im} \Pi^{(e)} (q^2) &=&  
-{\left \langle G^3
\right \rangle \over 5!\ 4!\ 2^{13} \pi^8} 
\int^{\Lambda}_0 du~ 
\Bigg \{ \fslash{q} \left [
q^2 \left ( {5u \over 2}-1 \right ) 
(1-u)  -m_c^2 \left ( {3u \over 2} + 7 \right ) 
\right ] 
\nonumber \\
&&+6 m_c q^2 (2u-1) -2 m_c^3 {3u+1\over 1-u}  \Bigg \}
u
\left ( -u q^2 + {m_c^2 u\over 1-u} \right )\ .
\end{eqnarray}
where the upper limit of the integrations is given by $\Lambda=1-m_c^2 /q^2$.
The integrations can be done analytically but we skip the messy analytic
expressions.
For the charm-quark propagators with two and three gluons attached, we use
the momentum-space expressions given in Ref.~\cite{Reinders:1984sr}.
In the $\Theta^+ (1540)$ sum rule~\cite{SDO03}, 
$\langle \bar {s} g_s \sigma \cdot G s \rangle$
was the important contribution. 
Since this condensate in the charm sector is replaced to 
$\langle G^3 \rangle$ in
the heavy quark expansion~\cite{Bagan:zp}, we include $\langle G^3 \rangle$
only from the heavy quark.
The Wilson coefficients for light-quark condensates are found to vanish 
except for
the $\langle \bar{q} q\rangle^4 $ terms shown in Fig.~\ref{fig1}(c). 
One can check that, for small $m_c$, our OPE reproduces the corresponding
OPE in Ref.~\cite{SDO03}.
The first term yields corresponding terms in Ref.~\cite{SDO03}, when it is
truncated up to $O(m_c)$.  
The second term in the limit $m_c \rightarrow 0$ gives the same gluon
condensate as in Ref.~\cite{SDO03}.
The fourth term involves $\left \langle {\alpha_s \over \pi} G^2
\right \rangle {1\over m_c}$ in the limit $m_c \rightarrow 0$.
When this part is converted to the quark condensate in the heavy quark
expansion, we find precisely the same Wilson coefficient of quark condensate
in Ref.~\cite{SDO03}.
Note that, similarly as in the phenomenological side, the OPE   has a chiral
odd and even part,
\begin{eqnarray}
\Pi^{\rm ope} (q)=\Pi^{\rm ope}_1(q^2)+\fslash{q} \Pi^{\rm ope}_q(q^2)\ .
\end{eqnarray}

\section{QCD sum rules and analysis}

QCD sum rules for $\Theta_c$ are constructed by matching the two spectral
densities in the Borel-weighted integral,
\begin{eqnarray}
\int^{S_0}_{m_c^2} 
dq^2 e^{-q^2/M^2} {1\over \pi}{\rm Im} [ \Pi_i^{\rm phen} (q^2)
- \Pi_i^{\rm ope} (q^2)] =0\ , \qquad (i=1,q)\ ,
\end{eqnarray}
where $M^2$ is the Borel mass.
Here, higher resonance contributions are subtracted according to 
the QCD duality assumption, which introduces the continuum threshold 
$S_0$.
As the correlator contains the chiral odd and even part, we have
two sum rules correspondingly,
\begin{eqnarray}
|\lambda_{\pm}|^2 e^{-m_\Theta^2 /M^2}
&=&\int^{S_0}_{m_c^2}
dq^2 e^{-q^2/M^2} {1\over \pi}{\rm Im} [ \Pi_q^{\rm ope} (q^2)]
\ ,\label{even}
\\
\pm m_\Theta |\lambda_{\pm}|^2 e^{-m_\Theta^2 /M^2}
&=&\int^{S_0}_{m_c^2}
dq^2 e^{-q^2/M^2} {1\over \pi}{\rm Im} [ \Pi_1^{\rm ope} (q^2)]
\ .\label{odd}
\end{eqnarray}
The second equation shows that the parity of $\Theta_c$ can be determined by
the sign of its right-hand side (RHS).
We use the following QCD parameters in our sum rules~\cite{Shifman:bx},
\begin{eqnarray}
\left \langle {\alpha_s \over \pi} G^2
\right \rangle &=& (0.33~{\rm GeV})^4 \ , \quad
\langle G^3 \rangle = 0.045 ~{\rm GeV}^6\ ,
\nonumber \\
m_c &=& 1.26~{\rm GeV} \ , \quad
\langle \bar {q} q \rangle = -(0.23~{\rm GeV})^3\ .
\end{eqnarray}
The value for the quark condensate corresponds to $m_q =7$
MeV ($q=u,d$) in the Gell-Mann--Oakes--Renner relation.
The values for the quark and gluon condensates are rather standard in
baryon or meson sum rules. For the charm quark mass,
other values can be found in the literature,  
1.1 GeV for its running mass and  
1.5 GeV for the pole mass~\cite{Narison:2001pu}. These
fit the open charm mass $M_D$ in the
sum rule of pseudoscalar correlator.   
Even a larger value is reported in Ref. \cite{Eidemuller:2002wk}. 
Below, we will check the sensitivity of our result to this parameter
as well as to the others. 

\begin{figure}[t]
\centering
\epsfig{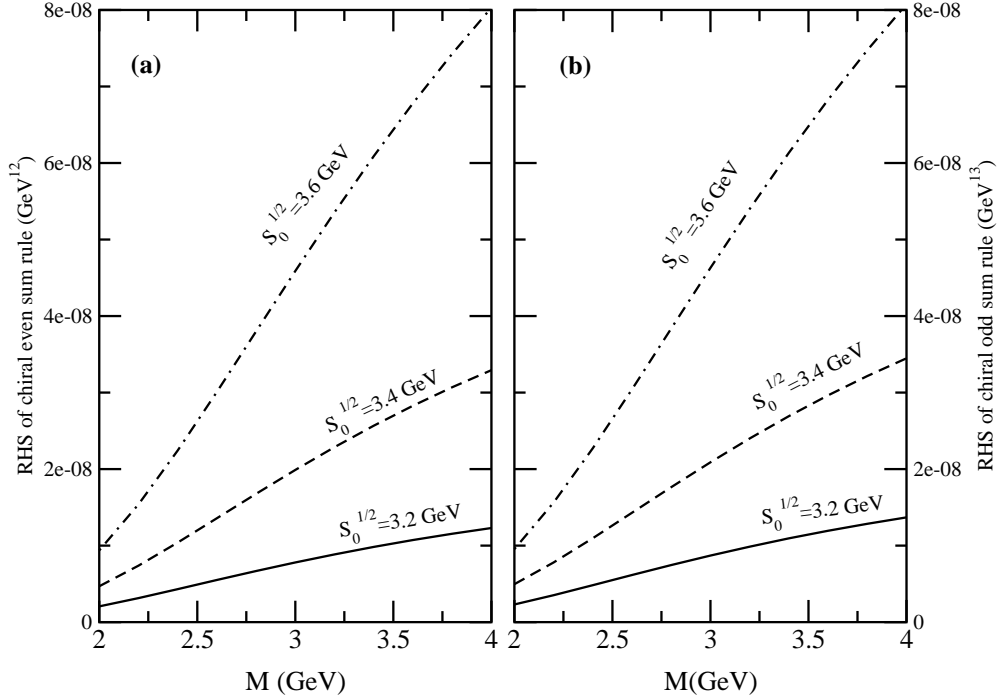}
\caption{(a) The RHS of chiral even sum rule of Eq. (\ref{even})
and (b) the RHS of the chiral odd sum rule Eq (\ref{odd}). 
The solid line is for $\sqrt{S_0}=3.2$ GeV, the dashed line for
$\sqrt{S_0}=3.4$ GeV and the dot-dashed line for $\sqrt{S_0}=3.6$ GeV.}
\label{fig2}
\end{figure}

In Fig.~\ref{fig2}, we plot the RHS of Eqs.~(\ref{even}) and (\ref{odd})
with respect to the Borel mass for various continuum thresholds      
$\sqrt{S_0} = 3.2$, $3.4$ and $3.6$ GeV.
As the left-hand side (LHS) of Eq.~(\ref{even}) is positive definite,
we check whether the RHS of Eq.~(\ref{even}) is consistently positive.
Our result in Fig.~\ref{fig2}(a) indeed shows that the RHS satisfies
this constraint.
In Fig.~\ref{fig2}(b), we plot the RHS of the chiral-odd sum rule (\ref{odd}).
Here, we see that the RHS is positive suggesting that the parity of
$\Theta_c$ is positive.
This positive parity agrees with the quenched lattice calculation
\cite{Chiu:2004uh}.
The two observations do not change as we vary the continuum threshold from
$\sqrt{S_0} = 3.2$ to 3.6 GeV.
The main contribution from the OPE is from the perturbative part 
[Fig.~\ref{fig1}(a)] and the gluon condensate coming from the light quark
[Fig.~\ref{fig1}(b)].
The contribution from $ \langle \bar {q} q \rangle $ and 
$\langle G^3 \rangle$ is found to be small and therefore the uncertainties
in these parameters
affect the result marginally.
Other values for $m_c$ ranging from 1.1 GeV to 1.5 GeV are found to yield
somewhat different curves but all of them have the same sign again
supporting the positive parity.

Our result favoring the positive parity is consistent with the
expectation from the constituent quark model picture with diquark
correlations by Jaffe and Wilczek~\cite{JW03}.
Namely, in the Jaffe-Wilczek picture, the two-diquark system has 
negative parity and, combining with the anticharm quark with
intrinsic negative parity, the pentaquark has the positive parity in total.
Note, in the SDO sum rule for $\Theta^+$ \cite{SDO03}, this picture
was not met by the large contribution from
the quark-gluon mixed condensate, 
$\langle \bar {s} g_s \sigma \cdot G s \rangle$, which is proportional
to the quark virtuality.  This mixed condensate is now
replaced by Fig.~\ref{fig1}(e) through the heavy quark expansion, whose
contribution to the OPE becomes marginal in our $\Theta_c$ sum rule.  
This constitutes the main mechanism for yielding positive parity 
in this heavy-light pentaquark system.
Therefore, we have an interesting crossover from the strange sector to
the charm sector in the pentaquark parity.

To check further the reliability of our sum rules, we calculate
the $\Theta_c$ mass and see if it agrees with the experimental value.
The $\Theta^+$ mass is determined in two ways.
\begin{enumerate}
\item
We take derivative of the chiral even sum rule (\ref{even}) with respect to
$1/M^2$ and divide the resulting equation by Eq. (\ref{even}).
As can be seen from Eq. (\ref{even}), this step leads to $-m^2_\Theta$ in
the LHS.
The mass is then obtained by taking negative square root of the resulting RHS.
\item
The same method as 1 applied to the chiral-odd sum rule, Eq. (\ref{odd}).
\end{enumerate}

\begin{figure}[t]
\centering
\epsfig{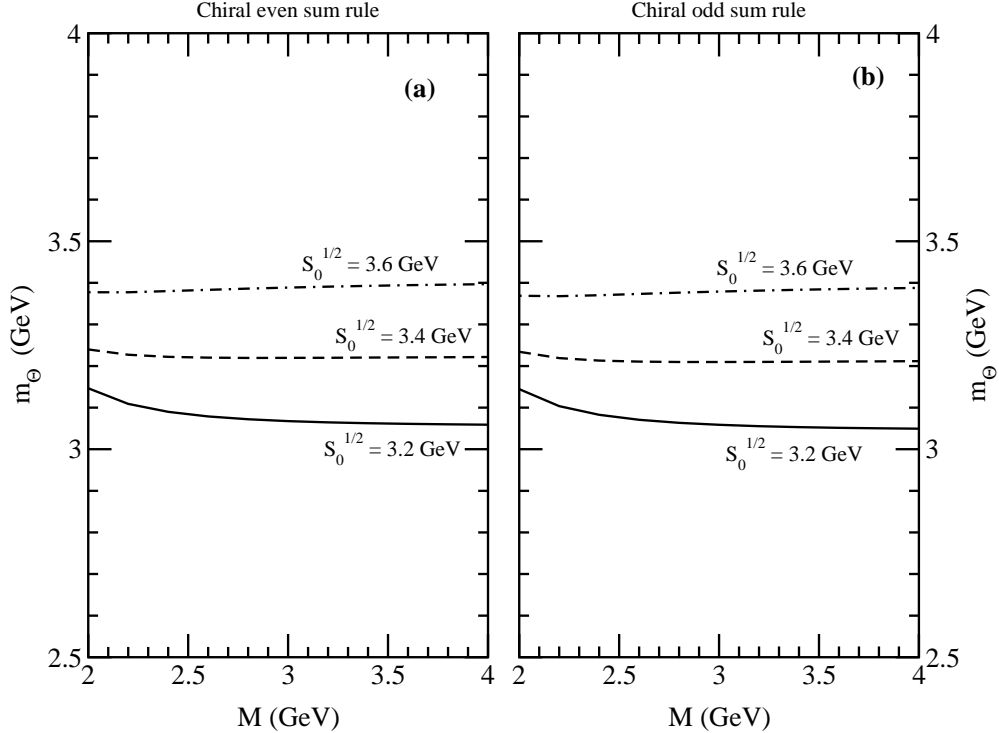}
\caption{The predicted $\Theta^+$ mass from chiral even sum rule (a) and
chiral odd sum rule (b) determined as explained in the text.
The notations are the same as Fig.~\ref{fig2}.}
\label{fig3}
\end{figure}

The results are shown in Fig.~\ref{fig3} for the two cases.
At the continuum threshold $\sqrt{S_0}=3.2$ GeV, both sum rules yield the
$\Theta_c$ mass to be around 3.06 GeV and the result is practically independent
of the Borel mass $M$. Changing the charm quark mass to a lower value of 
1.1 GeV leads to 3.02 GeV. For a much higher value of $m_c =1.5$ GeV,
we have 3.14 GeV at the stable Borel region.  So 
the extracted mass is slightly sensitive
to the charm quark mass. The extracted mass is more sensitive
to the continuum threshold. 
As we increase the continuum threshold, the result increases slightly also.
At the large threshold of $\sqrt{S_0}=3.6$ GeV, the predicted mass is around
3.4 GeV.  
Thus, our sum rules give the $\Theta_c$ mass that qualitatively agrees with
the experimental value of 3.099 GeV. 

To summarize, we have constructed QCD sum rules for the recently discovered
anti-charmed pentaquark $\Theta_c (3099)$.
The charm quark mass is kept finite in the OPE while for the light quark
propagators we use the coordinate space expressions obtained from the
expansion in the small quark mass.
Our sum rules suggest that the parity of $\Theta_c$ is positive, which
is opposite to that of $\Theta^+ (1540)$ determined
from QCD sum rules.
The obtained mass is qualitatively consistent with the experimental value
of the H1 Collaboration.

\acknowledgments

We are grateful to M. Oka and Fl. Stancu for fruitful discussions.
The work of S.H.L was supported by KOSEF under Grant No. 1999-2-111-005-5.
The work of Y.O. was supported by Forschungszentrum-J{\"u}lich, contract
No. 41445282 (COSY-058).


\begin{thebibliography}{10}

\bibitem{LEPS03}
\mbox{LEPS Collaboration,} T.~Nakano {\em et~al.\/},
  Phys. Rev. Lett. 91 (2003) 012002.

\bibitem{OTHERS}
S.~Stepanyan {\it et al.}  [CLAS Collaboration],
  Phys. Rev. Lett. 91 (2003) 252001;
J.~Barth {\em et~al.\/} [SAPHIR Collaboration],
  Phys. Lett. B 572 (2003) 127;
V.~V.~Barmin {\it et al.}  [DIANA Collaboration],
  Phys.\ Atom.\ Nucl.\  66 (2003) 1715
  [Yad.\ Fiz.\  66 (2003) 1763];
V.~Kubarovsky, S.~Stepanyan  [CLAS Collaboration],
  AIP Conf.\ Proc.\  698 (2004) 543;
A.~E. Asratyan, A.~G. Dolgolenko, M.~A. Kubantsev,
  hep-ex/0309042;
V.~Kubarovsky {\em et~al.\/} [CLAS Collaboration],
  Phys. Rev. Lett. 92 (2004) 032001;
A.~Airapetian, et~al. [HERMES Collaboration],
  Phys. Lett. B 585 (2004) 213;
S.~Armstrong, B.~Mellado and S.~L.~Wu,
  hep-ph/0312344;
A.~Aleev, et~al. [SVD Collaboration],
  hep-ex/0401024.

\bibitem{Alt:2003vb}
C.~Alt {\it et al.} [NA49 Collaboration],
Phys.\ Rev.\ Lett.\  92 (2004) 042003.

\bibitem{Aktas:2004qf}
A.~Aktas {\it et al.} [H1 Collaboration],
hep-ex/0403017.

\bibitem{Lip87-GSR87}
C.~Gignoux, B.~Silvestre-Brac, J.~M. Richard,
  Phys. Lett. B 193 (1987) 323;
H.~J. Lipkin,
  Phys. Lett. B 195 (1987) 484.

\bibitem{Stan98}
Fl. Stancu,
  Phys. Rev. D 58 (1998) 111501;
M. Genovese, J.-M. Richard, Fl. Stancu, S. Pepin,
  Phys. Lett. B 425 (1998) 171.

\bibitem{RS93}
D.~O. Riska, N.~N. Scoccola,
  Phys. Lett. B 299 (1993) 338.

\bibitem{OPM94}
Y.~Oh, B.-Y. Park, D.-P. Min,
  Phys. Lett. B 331 (1994) 362;
  Phys. Rev. D 50 (1994) 3350;
Y.~Oh, B.-Y. Park,
  Phys. Rev. D 51 (1995) 5016.

\bibitem{CBSC}
K.~Cheung,
  hep-ph/0308176;
T.~E. Browder, I.~R. Klebanov, D.~R. Marlow,
  Phys. Lett. B 587 (2004) 62;
I.~W. Stewart, M.~E. Wessling, M.~B. Wise,
  hep-ph/0402076;
M.~A.~Nowak, M.~Praszalowicz, M.~Sadzikowski, J.~Wasiluk,
  hep-ph/0403184;
H.~Y. Cheng, C.~K. Chua, C.~W. Hwang,
  hep-ph/0403232.

\bibitem{JW03}
R.~L.~Jaffe and F.~Wilczek,
Phys.\ Rev.\ Lett.\  91 (2003) 232003.

\bibitem{KL03a}
M.~Karliner, H.~J. Lipkin,
  hep-ph/0307243.

\bibitem{Kochelev:2004nd}
N.~I.~Kochelev, H.~J.~Lee, V.~Vento,
  hep-ph/0404065.

\bibitem{DPP97}
D.~Diakonov, V.~Petrov, M.~Polyakov,
  Z. Phys. A 359 (1997) 305.

\bibitem{CCKN03b}
C.~E.~Carlson, C.~D.~Carone, H.~J.~Kwee, V.~Nazaryan,
  Phys.\ Lett.\ B 579 (2004) 52.

\bibitem{Chiu:2004gg}
T.~W.~Chiu, T.~H.~Hsieh,
  hep-ph/0403020.

\bibitem{SDO03}
J.~Sugiyama, T.~Doi, M.~Oka,
  Phys.\ Lett.\ B 581 (2004) 167.

\bibitem{Sasaki03}
S.~Sasaki,
  hep-lat/0310014.

\bibitem{CFKK03}
F.~Csikor, Z.~Fodor, S.~D.~Katz, T.~G.~Kovacs,
  JHEP 0311 (2003) 070.

\bibitem{Oh:2003kw}
Y.~Oh, H.~Kim, S.~H.~Lee,
  Phys.\ Rev.\ D 69 (2004) 014009.

\bibitem{HHO03}
T.~Hyodo, A.~Hosaka, E.~Oset,
  Phys.\ Lett.\ B 579 (2004) 290.

\bibitem{ZANA}
Q.~Zhao,
  Phys.\ Rev.\ D 69 (2004) 053009;
K.~Nakayama, W.~G.~Love,
  hep-ph/0404011.

\bibitem{NT03}
K.~Nakayama, K.~Tsushima,
  Phys.\ Lett.\ B 583 (2004) 269.

\bibitem{ppst}
A.~W.~Thomas, K.~Hicks, A.~Hosaka,
  Prog.\ Theor.\ Phys.\  111 (2004) 291;
C. Hanhart {\it et al.\/},
  hep-ph/0312236;
S.~I.~Nam, A.~Hosaka, H.~C.~Kim,
  hep-ph/0401074.

\bibitem{Oh:2003xg}
Y.~Oh, H.~Kim, S.~H.~Lee,
  hep-ph/0312229.

\bibitem{Zhu03}
S.~L.~Zhu,
  Phys.\ Rev.\ Lett.\  91 (2003) 232002.

\bibitem{MNNRL03}
R.~D.~Matheus, F.~S.~Navarra, M.~Nielsen, R.~Rodrigues da Silva, S.~H.~Lee,
  Phys.\ Lett.\ B 578 (2004) 323.

\bibitem{Eidemuller:2004ra}
M.~Eidemuller,
  hep-ph/0404126.

\bibitem{Chiu:2004uh}
T.~W.~Chiu, T.~H.~Hsieh,
  hep-ph/0404007.

\bibitem{Griegel:rn}
D.~K.~Griegel,
{\em Nucleon Propagation In Nuclear Matter: A QCD Sum Rule Approach},
Ph.D Thesis, Univeristy of Maryland, 1991.

\bibitem{Jido:1996ia}
D.~Jido, N.~Kodama, M.~Oka,
  Phys.\ Rev.\ D 54 (1996) 4532.

\bibitem{Lee:2002jb}
F.~X.~Lee, X.~Y.~Liu,
  Phys.\ Rev.\ D 66 (2002) 014014.

\bibitem{Jin:1997pb}
X.~M.~Jin, J.~Tang,
  Phys.\ Rev.\ D 56 (1997) 515.

\bibitem{Reinders:1984sr}
L.~J.~Reinders, H.~Rubinstein, S.~Yazaki,
  Phys.\ Rept.\  127 (1985) 1.

\bibitem{Bagan:zp}
E.~Bagan, J.~I.~Latorre, P.~Pascual,
  Z.\ Phys.\ C 32 (1986) 43.

\bibitem{Shifman:bx}
M.~A.~Shifman, A.~I.~Vainshtein, V.~I.~Zakharov,
  Nucl.\ Phys.\ B 147 (1979) 385.

\bibitem{Narison:2001pu}
S.~Narison,
  Phys.\ Lett.\ B 520 (2001) 115.

\bibitem{Eidemuller:2002wk}
M.~Eidemuller,
  Phys.\ Rev.\ D 67 (2003) 113002.


\end{thebibliography}
\end{document}